# Quantized Ballistic Transport of Electrons and Electron Pairs in LaAlO$_3$/SrTiO$_3$ Nanowires

Anil Annadi[1,2], Guanglei Cheng[1,2,4], Hyungwoo Lee[3], Jung-Woo Lee[3], Shicheng Lu[1,2], Anthony Tylan-Tyler[1,2], Megan Briggeman[1,2], Michelle Tomczyk[1,2], Mengchen Huang[1,2], David Pekker[1,2], Chang-Beom Eom[3], Patrick Irvin[1,2], Jeremy Levy[1,2*]

[1]Department of Physics and Astronomy, University of Pittsburgh, Pittsburgh, PA 15260, USA.

[2]Pittsburgh Quantum Institute, Pittsburgh, PA, 15260 USA.

[3]Department of Materials Science and Engineering, University of Wisconsin-Madison, Madison, WI 53706, USA.

[4]CAS Key Laboratory of Microscale Magnetic Resonance and Department of Modern Physics, University of Science and Technology of China, Hefei 230026, China

* jlevy@pitt.edu

**Abstract:**

SrTiO$_3$-based heterointerfaces support quasi-two-dimensional (2D) electron systems that are analogous to III-V semiconductor heterostructures, but also possess superconducting, magnetic, spintronic, ferroelectric, and ferroelastic degrees of freedom. Despite these rich properties, the relatively low mobilities of 2D complex-oxide interfaces appear to preclude ballistic transport in 1D. Here we show that the 2D LaAlO$_3$/SrTiO$_3$ interface can support quantized ballistic transport of electrons and (non-superconducting) electron pairs within quasi-1D structures that are created using a well-established conductive atomic-force microscope (c-AFM) lithography technique. The nature of transport ranges from truly single-mode (1D) to three-dimensional (3D), depending on the applied magnetic field and gate voltage. Quantization of the lowest $e^2/h$ plateau indicate a ballistic mean-free path $l_{MF} \sim 20$ μm, more than two orders of magnitude larger than for 2D LaAlO$_3$/SrTiO$_3$ heterostructures. Non-superconducting electron pairs are found to be stable in magnetic fields as high as $B = 11$ T, and propagate ballistically with conductance quantized at $2e^2/h$. Theories of one-dimensional (1D) transport of interacting electron systems depend crucially on the sign of the electron-electron interaction, which may help explain the highly ballistic transport behavior. The 1D geometry yields new insights into the electronic structure of the LaAlO$_3$/SrTiO$_3$ system and offers a new platform for the study of strongly interacting 1D electronic systems.

Dimensionality has a profound effect on electron transport. When electrons are confined in two dimensions (2D), new phases such as the integer[1] and fractional[2] quantum Hall effect emerge. Electrons confined in one dimension (1D) lose nearly all of their recognizable features[3-4]. For example, the electron spin and charge can separate and move independently of one another[5], and the charge itself can fractionalize (Ref. 6). However, in 1D, the conductance remains quantized in units of $e^2/h$ (Ref. 7). The edges of 2D quantum Hall systems form nearly-ideal 1D channels, where magnetic confinement gaps out the 2D bulk and protects electrons from back-scattering. The chiral edge transport of the quantum Hall phase is fundamentally different from transport in 1D nanostructures where electrons are electrostatically confined to a narrow channel. Quasi-1D transport was first reported in narrow constrictions, also known as "quantum point contacts"[8-9]. The conductance through these narrow channels is given by the number of allowed transverse modes, which is tunable by an external gate. The confined regions are generally short, of the order 100-200 nm, with a channel length set by the distance between the top gate electrodes and the high-mobility buried layer. There have been various attempts to engineer more extended 1D quantum wires using other growth techniques and different materials. For example, cleaved-edge overgrown III-V quantum wires exhibit quantized transport in devices[10]. Other one-dimensional systems include carbon nanotubes[11], graphene nanoribbons[12], and compound semiconductor nanowires[10, 13]. In all of these systems, electron transport is sensitive to minute amounts of disorder. For example, when 2D semiconductor heterostructures are patterned into 1D channels, the mobility drops significantly[14]. Theoretically, this sensitivity to disorder can be understood within the framework of Tomonaga-Luttinger liquid theory, which predicts that repulsive interactions promote full backscattering from even a single weak impurity[15-16]. Conversely, attractive interactions are predicted to strongly suppress impurity scattering[16-17].

Oxide heterostructures have added new richness to the field of quantum transport in the last decade. For example, ZnO/(Mn,Zn)O heterostructures have achieved sufficiently high mobility to reveal fractional quantum Hall states[18], which has revealed new even-denominator states not visible in III-V hosts[19]. LaAlO$_3$/SrTiO$_3$ heterostructures[20] exhibit a wide range of behavior including gate-tunable conducting[21], superconducting[22], ferromagnetic[23], and spin-orbit coupled[24-27] phases. As interesting and rich as its palette of phases may be, the 2D electron mobility is still low ($\mu_H \sim 10^3$ cm$^2$/Vs) compared with high-mobility GaAs/AlGaAs heterointerfaces ($\mu_H \sim 10^7$ cm$^2$/Vs). However, despite the modest mobility of the LaAlO$_3$/SrTiO$_3$ 2D interface, there is an increasing body of evidence suggesting that 1D geometries are able to support ballistic transport[24-27].

Transport through a coherent quantum conductor can be described by Landauer's formula, $G = (e^2/h) \sum_i T_i(\mu)$, where each energy subband available at chemical potential $\mu$ contributes one quantum of conductance $e^2/h$ with transmission probability $T_i(\mu)$. The transmission probability is given by $T_i(\mu) = \bar{T} F_T(\mu - E_i)$ where $\bar{T}$ encompasses any tunneling resonances, cavity interference effects, or backscattering processes, $F_T(E)$ is a thermal broadening from the Fermi distribution function of the leads at a finite temperature, and $E_i$ represents the energy minimum of the $i^{\text{th}}$ electron subband[28]. For simplicity, we assume that $\bar{T}$ is independent of energy. Within this framework, the conductance increases in steps of $e^2/h$ every time the chemical potential crosses a subband energy minimum. Transport through the channel is ballistic and dissipationless; however, the measured resistance is given by $R = h/(Ne^2)$, where $N$ is the number of occupied subbands. The apparent contradiction between dissipationless transport within the waveguide and finite resistance was understood by Landauer, and put on a rigorous footing by Maslov and Stone, who developed a Luttinger liquid model of energy dissipation within the leads[29]. However, in experiments, even the cleanest non-chiral systems do not have infinite scattering lengths; each subband can backscatter electrons, leading to a suppression which can be modeled as $\bar{T} = \exp(-L/L_i)$ (Ref. 30), where $L$ is the channel length and $L_i$ is the mode-dependent scattering length. When $L_i \sim L$, the system is in the ballistic or quasi-ballistic regime, and when $L_i \gg L$, the system enters a quantized ballistic regime.

The expected properties of an ideal few-mode (i.e., few-subband) electron waveguide are illustrated in Figure 1. The conductance of the waveguide depends on the number of accessible quantum channels (shown in Figure 1D-E as energy-shifted parabolic bands), which is controlled by the applied side-gate voltage of the device $V_{sg}$. Figure 1B, D depicts a state in which a single spin-resolved subband is occupied. As the chemical potential $\mu$ is increased, more subbands in the waveguide become occupied. Figure 1C and Figure 1E depict a state in which $N = 3$ subbands contribute to transport. Each spin-resolved subband contributes $e^2/h$ to the total conductance (Figure 1F). The energy at which $\mu$ crosses a new subband (at $k_x = 0$) can generally shift in an applied magnetic field due to Zeeman and orbital effects. When lateral and vertical confinement energies are comparable, a more complex subband structure can emerge, as illustrated in Figure 1G.

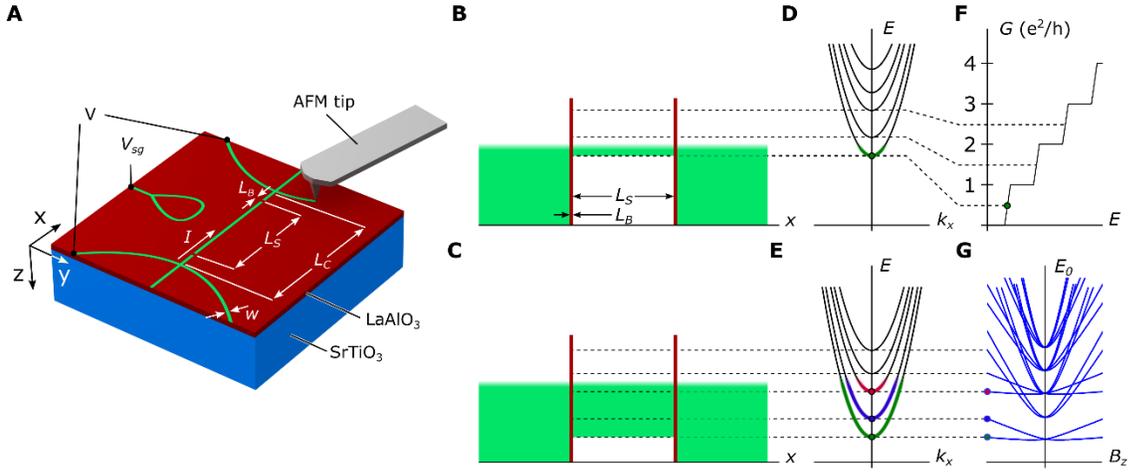

**Figure 1. Expected transport characteristics of electron waveguides.** (**A**) Schematic of LaAlO$_3$/SrTiO$_3$ electron waveguides fabricated by c-AFM lithography. Two barriers, with width $L_B \sim 5-10$ nm and spacing $L_S \sim 10-1000$ nm, formed by negative AFM tip pulses, surround the main channel of length $L_C \sim 500-1800$ nm, enabling the subbands in the waveguide to be tuned by $V_{sg}$. (**B** and **C**) Energy diagrams of the waveguide for two different values of chemical potential $\mu$, which is controlled by $V_{sg}$. For (**B**), a single subband is occupied, while for (**C**) three subbands are occupied. (**D** and **E**) depict the corresponding energy subbands corresponding to (**B**) and (**C**). Thick colored bands indicate occupied states. (**F**) Zero-bias conductance quantization as a function of chemical potential. (**G**) Waveguide subband structure (with both lateral and vertical confinement) as a function of magnetic field (which couples to the electrons via both Zeeman and orbital effect) and chemical potential.

LaAlO$_3$/SrTiO$_3$ samples are grown by pulsed laser deposition (PLD) under conditions that are described in detail elsewhere[31]. The electron waveguides are created using c-AFM lithography technique[32-33]. Positive voltages applied between the c-AFM tip and the top LaAlO$_3$ surface locally produce conductive regions at the LaAlO$_3$/SrTiO$_3$ interface (illustrated in Figure 1A), while negative voltages locally restore the insulating phase. The mechanism for writing (erasing) is attributed to LaAlO$_3$ surface protonation (de-protonation)[34-35]. The protonated LaAlO$_3$ surface in critical-thickness LaAlO$_3$/SrTiO$_3$ heterostructures creates an attractive confining potential that defines the nanowire. Because the protons are physically separated from the conducting region by a highly insulating LaAlO$_3$ barrier, this nanofabrication method can be viewed as analogous to the "modulation doping" technique[36] commonly used in III-V semiconductor heterostructures. The separation of dopants from the conducting region minimizes scattering from imperfections. A key difference from III-V nanostructures is the relative proximity between the dopant layer and conducting channel, here only 1.2 nm. Typical nanowire widths at room temperature are $w$~10 nm, as measured by local erasure experiments[32].

We fabricate LaAlO$_3$/SrTiO$_3$ electron waveguides using a well-established conductive atomic-force microscopy (c-AFM) lithography technique[32-33], as shown in Figure 1A (also see Materials and Methods). The waveguide geometry consists of a nanowire channel of total length $L_C$, surrounded by two narrow, highly transparent barriers (width $L_B$~5 −20 nm) separated by a distance $L_S \sim 10 - 1000$ nm. The experimentally measured conductance of LaAlO$_3$/SrTiO$_3$ waveguides is shown in Figure 2A-D. We focus on two distinct devices: device A ($L_C = 500$ nm, $L_S = 50$ nm, $L_B = 20$ nm) and device B ($L_C = 1.8$ µm, $L_S = 1$ µm, $L_B = 20$ nm). Figure 2A and Figure 2C show the zero-bias conductance $G = dI/dV$ as a function of side-gate voltage $V_{sg}$ (or chemical potential $\mu$) for a sequence of magnetic fields between $B = 0$ T and 9 T. Analysis of the non-equilibrium conductance, described in Section S1 of the Supporting Information, enables the lever-arm ratio $\alpha \equiv d\mu/dV_{sg}$ and g-factor $g \equiv \mu_B^{-1} d\mu/dB$, where $\mu_B$ is the Bohr magneton for the two devices A (B), to be determined: $\alpha_{A(B)} = 4.5 \pm 0.2$ ($9.9 \pm 1.7$) µeV/mV and $g_{A(B)} = 0.62 \pm 0.03$ ($0.61 \pm 0.04$). For Device A (Figure 2A), clear conductance steps of $G = 2e^2/h$ are visible for magnetic fields above ~1 T. These steps split into $e^2/h$ steps, up to $N = 6$, at fields above ~3 T. These electron waveguides exhibit no valley degeneracies and can be tuned to the lowest spin-polarized conduction plateau ($G = e^2/h$), with no signatures of sub-structure or "0.7 anomalies" [37]. When only a single barrier is present no conduction quantization is observed (see discussion in Supporting Information and Figure S6B). When no barriers are present the overall conductance is very large and cannot be tuned to an insulating phase while maintaining the conductance of the voltage leads (Figure S6A).

We attribute the observed conduction plateaus to Landauer quantization[7], for which the total conductance depends on the number of available quantum channels (subbands). The subband structure of these LaAlO$_3$/SrTiO$_3$ electron waveguides is clearly revealed by examining the transconductance $dG/d\mu$ as a function of $\mu$ and external magnetic field $B$ (Figure 2B, D). The transconductance peaks (bright areas) mark the boundaries where new subbands become available (as illustrated in Figure 1G). The subbands are separated by regions where the conductance is highly quantized ($dG/d\mu \to 0$). At low magnetic fields (and low $\mu$), the subbands scale roughly as $B^2$ and become more linear at larger magnetic fields. A pattern of subbands repeats at least twice, spaced by approximately 500 μeV. The transconductance of the two devices A ($L_S = 50$ nm) and B ($L_S = 1$ μm) are remarkably similar, despite the large difference in channel length and the fact that the lever arm for the two devices differs by a factor of two.

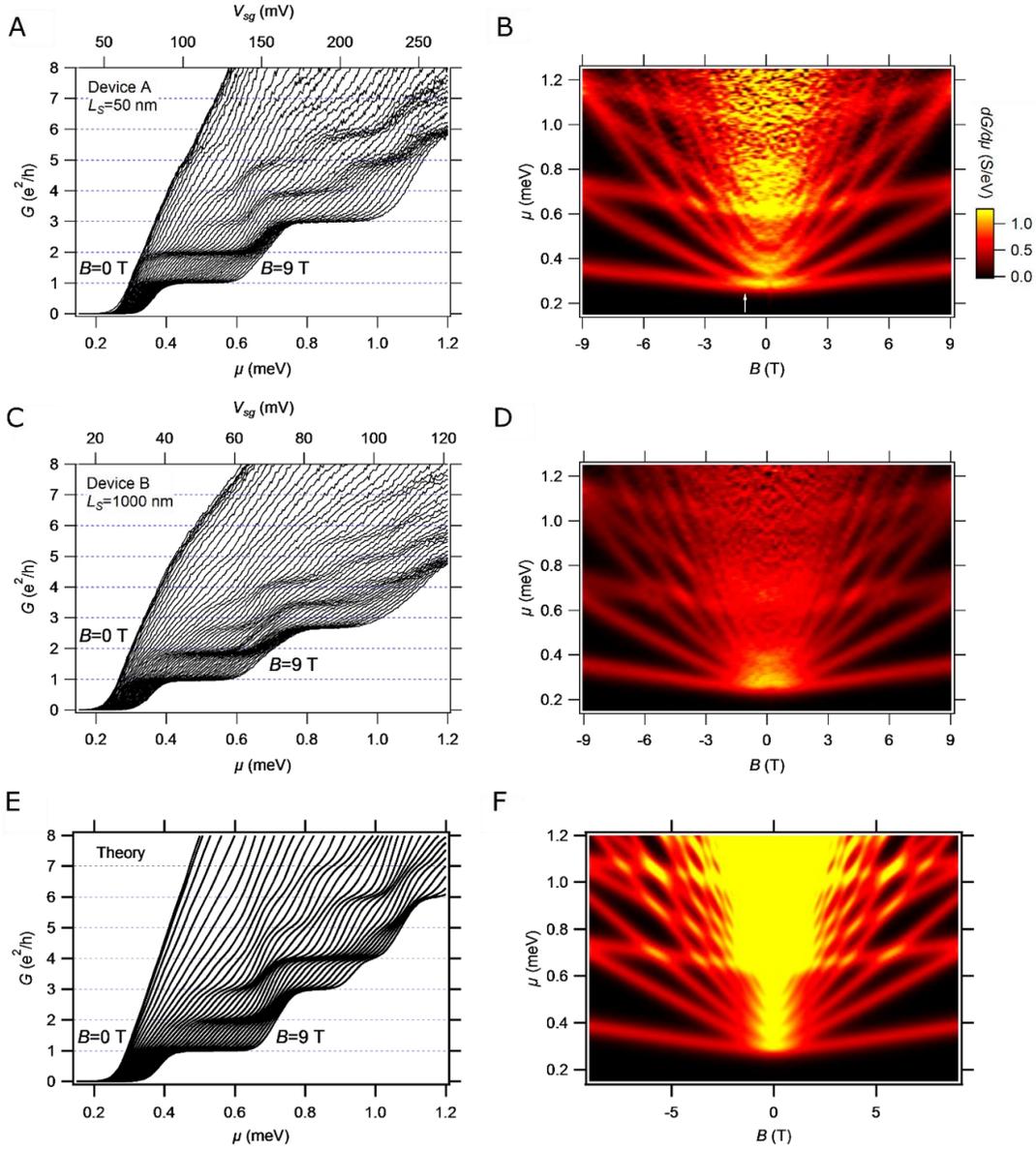

**Figure 2. Conductance and transconductance of devices A and B, and comparison with theory**. (**A** and **C**) Zero-bias conductance of device A ($L_B = 20$ nm, $L_C = 500$ nm, $L_S = 50$ nm) and device B ($L_B = 20$ nm, $L_C = 1,800$ nm, $L_S = 1,000$ nm) as a function of chemical potential $\mu$ and magnetic fields $B$ in the range 0-9 T, at T = 50 mK. (**B** and **D**) Transconductance $dG/d\mu$ shown as a function of $\mu$ and $B$ for device A (B) and device B (D). Each bright band marks the crossing of a subband, and the white arrow in (**B**) indicates the pairing field $B_P \approx 1$ T. The revealed subband structures show remarkable similarity between these two devices. (**E**) Theoretical zero-bias conductance curves, modeling device A, for a non-interacting channel as a function of the chemical potential and magnetic field. (**F**) Corresponding transconductance $dG/d\mu$ as a function of $\mu$ and $B$. Transitions have been broadened by a 65 µeV-wide Lorentzian. Experimental data is obtained at temperature $T = 50$ mK.

While the lowest $N = 1$ state remains highly quantized for both devices (see Figure 2), the plateaus do not fully reach the integer values for higher $N$ for device B. The relationship between two length scales—the length scale of the device and the elastic scattering length (which is typically much shorter than the inelastic scattering length in quantum devices) — determines whether transport is ballistic. The conductance of these modes are not exactly $e^2/h$, however, in part because they are not topologically-protected edge modes, nor are they quantum Hall edge states[38]. In electron waveguides at the LaAlO3/SrTiO3 interface, the elastic scattering length can be estimated by assuming an exponential decay of the conductance $G = G_0\exp(-L/L_0)$, where $L_0$ is the scattering length and $L$ is the length of the device. The location of the minimum in the transconductance is used to find the value of the plateaus, as seen in Figure S2. The scattering lengths greatly exceed the length of the devices (Table S1), implying that the transport is fully ballistic. The error estimate for Device A is limited by the short length of the channel. For Device B, the channel length is long enough to yield (with 10% accuracy) a measure of the scattering length, which is surprising given how low the 2D mobility is for LaAlO3/SrTiO3. We also note that systematic errors (e.g., reflections of incident electrons at one or both of the barriers) are only expected to increase these estimates.

Many of the features in the transconductance spectra shown in Figure 2A-D are captured by a waveguide model of non-interacting electrons in a 3D waveguide. The waveguide's confining potential can be regarded as translationally invariant along the propagation direction ($x$) and convex along the two transverse directions (lateral $y$ and vertical $z$). Since the measured carrier density in conductive nanostructures created by c-AFM lithography is typically in the range of $0.5 - 1.0 \times 10^{13}$ cm$^{-2}$ (Ref. 39), only the Ti $d_{xy}$ band, being lower in energy than the $d_{xz}$ and $d_{yz}$ bands at the LaAlO3/SrTiO3 interface[40], is expected to be occupied at these carrier densities. Thus we assume that all of the conducting channels are derived from the lower $d_{xy}$ band.

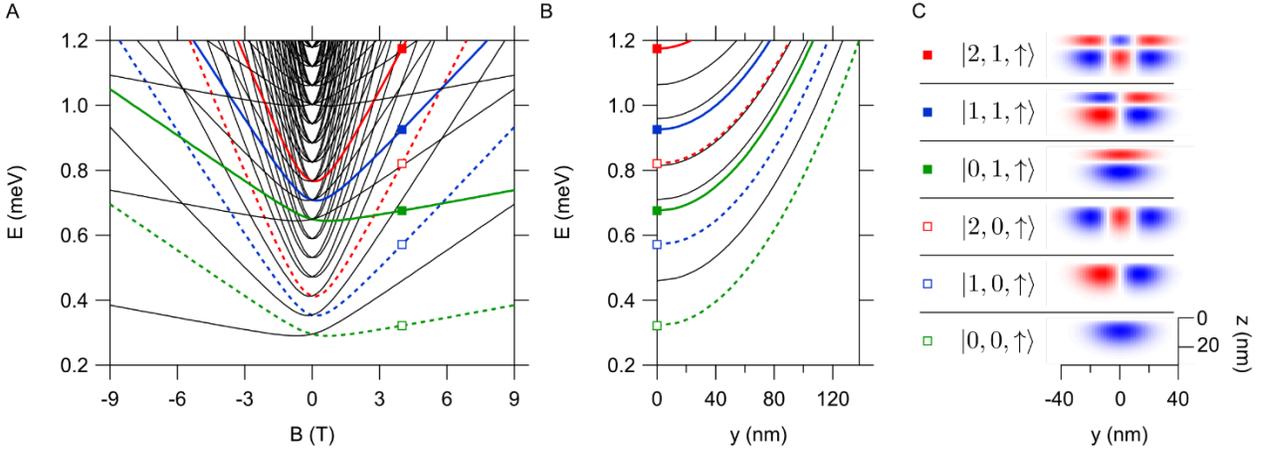

**Figure 3. Non-interacting waveguide model.** (**A**) Eigenenergies for a quantum wire for the Hamiltonian described in Eq. (1 are plotted as a function of magnetic field $B$. Selected spin-up states are highlighted in color. (**B**) Magnetically-induced displacement of these states along the $y$ direction as a function of eigenstate energy for $B = 4\,\text{T}$. (**C**) Six corresponding wavefunctions, labeled by $|n_y, n_z, s\rangle$, at $k_x = 0$ and $B = 4$ T. Red and blue colors indicate opposite sign of the wavefunction.

We use a potential $U_y = \frac{1}{2}m_y^*\omega_y^2 y^2$ to describe the lateral confinement, where $m_x^* = m_y^*$ is the effective mass in the x-y plane and $\omega_y = \frac{\hbar}{m_y^* \ell_y^2}$ is the confinement frequency with $l_y$ being the characteristic width of the waveguide. In the vertical direction, the confinement at the interface is modeled by a half-parabolic potential, namely, $U_z = \frac{1}{2}m_z^*\omega_z^2 z^2$ for $z > 0$ and $U_z = +\infty$ for $z \le 0$, where $\omega_z = \frac{\hbar}{m_z^* \ell_z^2}$ is the confinement frequency, $m_z^*$ is the effective mass of the $d_{xy}$ band in the z direction, and $l_z$ is the penetration depth into the SrTiO3. Within this single-particle picture, the full Hamiltonian can be written in the Landau gauge as

$$H = \frac{(p_x - eBy)^2}{2m_x^*} + \frac{p_y^2}{2m_y^*} + \frac{p_z^2}{2m_z^*} + \frac{m_y^*\omega_y^2}{2}y^2 + \frac{m_z^*\omega_z^2}{2}z^2 - g\frac{\mu_B}{2}B\sigma_z, \quad (1)$$

where $\sigma_z$ is the Pauli matrix. This Hamiltonian is readily solved to yield energy eigenstates $|n_y, n_z, s\rangle \otimes |k_x\rangle$ with corresponding energy

$$E_{m,n,s,k_x} = \hbar\Omega\left(n_y + \frac{1}{2}\right) + \hbar\omega_z\left((2n_z + 1) + \frac{1}{2}\right) - g\mu_B B s + \frac{\hbar^2 k_x^2}{2m_x^*}\left(1 - \frac{\omega_c^2}{\Omega^2}\right), \quad (2)$$

where $\omega_c = \frac{eB}{m_y^*}$ is the cyclotron frequency, $\Omega = \sqrt{\omega_y^2 + \omega_c^2}$ is the effective frequency of the waveguide and magnetic field, $n_y$ ($n_z$) enumerates the lateral (vertical) states, and $s = \pm 1/2$ is the spin quantum number. Distinct spin-resolved subbands[41] are associated with the discrete quantum numbers $|n_y, n_z, s\rangle$. Figure 3A plots the eigenenergies for parameters that have been adjusted to resemble the experimentally measured transconductance (Figure 2D). These values are also used to compute the expected conductance and transconductance versus chemical potential (Figure 2E, F). The corresponding wavefunctions $\phi_{n_y,n_z,k_x,s}(y,z)$ (defined in Eq. (3) for selected states are illustrated in Figure 3C:

$$\phi_{n_y,n_z,k,s}(y,z) \equiv \langle y,z,s; k|n_y, n_z, s\rangle \otimes |k_x\rangle$$

$$= N_{n_y,n_z,k} e^{-\frac{m_y^*\Omega}{2\hbar}\left(y - \frac{\hbar\omega_c^2}{m_y^*\Omega}k\right)^2} H_n\left(\sqrt{\frac{m_y^*\Omega}{\hbar}}\left(y - \frac{\hbar\omega_c^2}{m_y^*\Omega}k\right)\right) e^{-\frac{m_z^*\omega_z}{2\hbar}z^2} H_{2m+1}\left(\sqrt{\frac{m_z^*\omega_z}{\hbar}}z\right). \quad (3)$$

Here, $H_n(x)$ are the Hermite polynomials. The wavefunctions are displaced laterally by the magnetic field by an amount that depends quadratically on the kinetic energy (Figure 3B). The set of parameters for device A (B), $\ell_y = 26$ (27) nm, $\ell_z = 8.1$ (7.9) nm, $m_x^* = m_y^* = 1.9$

(1.8) $m_e$, and $m_z^* = 6.5$ (6.4) $m_e$ is obtained by maximizing agreement with a tight-binding model that includes spin-orbit interactions (see Supporting Information). At low magnetic fields, the energy scales quadratically with magnetic field, as it is dominated by the geometrical confinement contribution; at higher magnetic fields, the confinement from the cyclotron orbits dominates, producing a linear scaling. The crossover occurs near $\omega_B = \frac{eB}{m_y^*} \sim \omega_y$.

While the single-particle model captures the overall subband structure, clear deviations in the experimental results are apparent. For example, the lowest two subband minima device A and B ($|0,0,\uparrow\rangle$ and $|0,0,\downarrow\rangle$) merge not at zero magnetic field but at a critical field $B_p \approx 1$ T (see Figure 2). In other devices, this phenomenon is even more pronounced. Device C, written on a different sample, exhibits highly quantized conduction but with a subband structure that differs qualitatively from devices A and B. There are three pairs of subbands that generate $2e^2/h$ steps (Figure 4A). These states separate at a critical field $B_p \approx 11$ T (Figure 4C, dashed lines). Superimposed over these pairs is a separate subband (with higher curvature) that contributes $e^2/h$ to the conductance (Figure 4B). At $B \approx 3$ T two paired subbands are superimposed with the unpaired subband, leading to a plateau near $5e^2/h$ (highlighted in green).

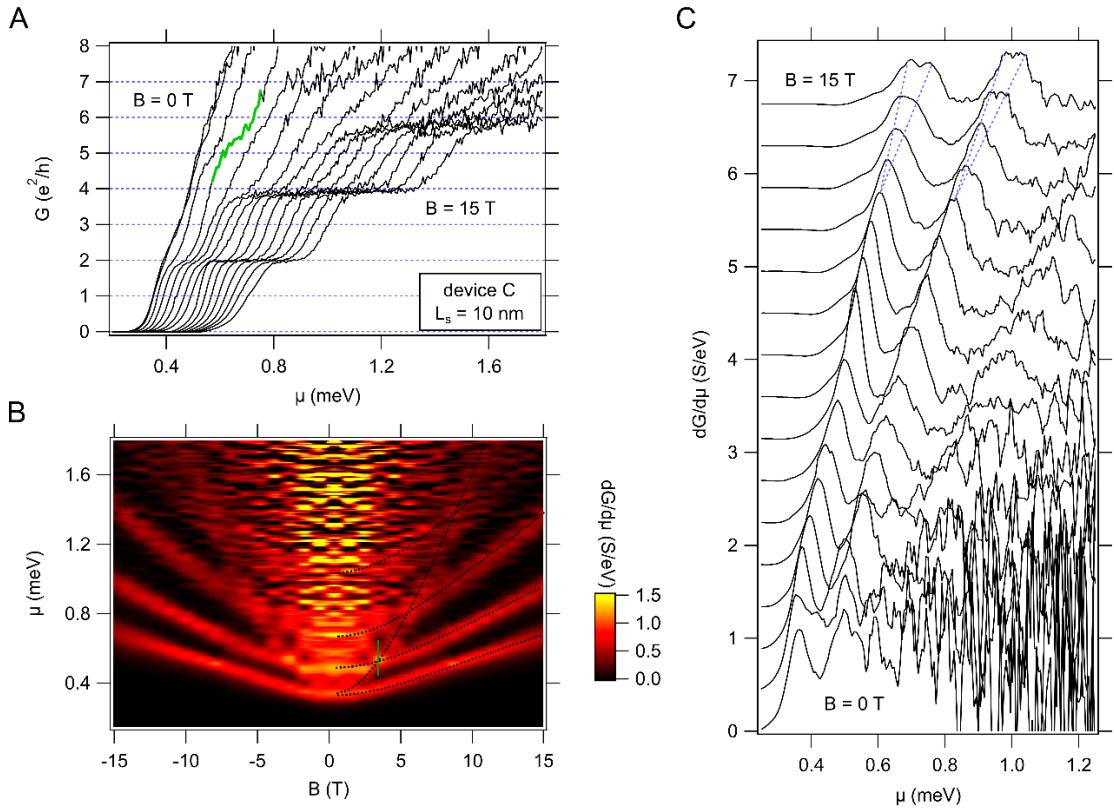

**Figure 4. Strongly paired electron waveguides.** (**A**) Conductance of device C ($L_B = 5$ nm, $L_C = 350$ nm, $L_S = 10$ nm) versus chemical potential for magnetic fields ranging from 0 T to 15 T, at T = 50 mK. This device shows strong electron pairing and associated $2e^2/h$ conductance steps. (**B**) Transconductance plot shows three strongly paired states and a superimposed state with higher curvature associated with a conductance of $e^2/h$. The value of the later state can be seen at $B = 3$ T in the conductance curve in panel a (highlighted in green) where it combines with the second strongly paired subband into a plateau near $5e^2/h$. (**C**) Linecuts of transconductance plotted at magnetic fields from 0 T to 15 T in 1 T steps. The $2e^2/h$ peaks split above a pairing field $B_p \approx 11$ T, as indicated by the dashed lines.

To explain this locking of subbands, we investigated a variety of single-particle interactions (e.g., band anisotropy, spin-orbit interactions). None of these interactions are able to reproduce this locking phenomena. This locking behavior can be accounted for by introducing attractive electron-electron interactions within the waveguide. Within this framework, locking of subbands is associated with a phase in which electrons are paired but not superconducting[42]. The effects of these interactions become apparent in the transconductance data in the vicinity of subband crossing points (both at zero magnetic field and at finite field). We a;sp observe extended regions of $2e^2/h$ conductance steps which we associate with a transition from a vacuum phase directly into a paired phase. That is, when a pair of subbands with opposite spin (e.g.$|1,0,\uparrow\rangle$ and $|0,1,\downarrow\rangle$) intersect at a finite magnetic field they are found to pair re-entrantly before separating again (Figure 5). This observation is consistent with previously studies of one dimensional fermions with attractive interactions using both the Bethe Ansatz approach[43] (for the case of equal masses) and numerical approaches[44-45] (for the case of unequal masses).

Here, we present a simple self-consistent Hartree-Bogoliubov model of crossing subbands that is both consistent with the more refined approaches and highlights the relevant physics without added complication. We start with the two-band, one-dimensional Hubbard model:

$$H = -\sum_{i,\alpha} t_\alpha (c^\dagger_{\alpha,i} c_{\alpha,i+1} + h.c.) + \sum_{i,\alpha} V_\alpha(V_{sg}, B)\, n_{\alpha,i} + U \sum_i n_{1,i} n_{2,i}, \quad (4)$$

where $i$ is the site index, $\alpha$ is the subband index, $V_\alpha(V_{sg}, B)$ describes the electrochemical potential as a function of the side gate voltage and magnetic field, and $U < 0$ models the electron-electron attraction. At the mean field level, this model is described by the single-particle Hamiltonian

$$\begin{pmatrix} \xi_{1,k} + \Sigma_1 & 0 & 0 & \Delta_{rp} \\ 0 & -(\xi_{1,k} + \Sigma_1) & \Delta_{rp} & 0 \\ 0 & \Delta_{rp} & \xi_{2,k} + \Sigma_2 & 0 \\ \Delta_{rp} & 0 & 0 & -(\xi_{2,k} + \Sigma_2) \end{pmatrix} \psi_{\beta,k} = E_{\beta,k} \psi_{\beta,k}, \quad (5)$$

where we use the $\{c_{1,k}, c_{1,k}^\dagger, c_{2,-k}, c_{2,-k}^\dagger\}$ basis, $\{1,2\}$ are the subband labels, $\psi_{\beta,k}$ and $E_{\beta,k}$ are the quasi-particle wave functions and eigenenergies, $\xi_{\alpha,k}(\mu, B)$ corresponds to the non-interacting energy of an electron in the transverse subband $\alpha$ with momentum $k$ along the wire, in magnetic field $B$, and chemical potential $\mu$ (that is tuned by $V_{sg}$). $\Sigma_1, \Sigma_2, \Delta_{rp}$ are the mean fields that must be found self-consistently. $\Sigma_\alpha$ represents the Hartree shifts due to the electrons in the opposite subband $\bar{\alpha}$:

$$\Sigma_\alpha = U_H \int \frac{dk}{2\pi} \langle c_{\bar{\alpha},k}^\dagger c_{\bar{\alpha},k} \rangle \tag{6}$$

and $\Delta_{rp}$ represents the re-entrant pairing field

$$\Delta_{rp} = U_B \int \frac{dk}{2\pi} \langle c_{2,-k} c_{1,k} \rangle. \tag{7}$$

For concreteness, we have made the minimal assumption that the interactions are momentum-independent (i.e. local in real space) when writing the mean fields. We caution that a nonzero value of $\Delta$ should not be interpreted as a signature of superconductivity but only as a signature of pair formation as we are working in one dimension. Finally, when computing the matrix elements, we must keep in mind that the basis we are using is twice as big as the physical basis and consequently, quasi-particle wave functions come in conjugate pairs. Only one member of the pair should be used (e.g. the one that has the positive eigenvalue and thus corresponds to the quasi-particle creation operator).

We solve the Hartree-Bogoliubov model self-consistently to obtain a phase diagram near the crossing point of the $|0,1,\downarrow\rangle$ and $|1,0,\uparrow\rangle$ subbands (Figure 5C). The locations of the non-interacting subbands are plotted with dashed lines. By turning on the attractive inter-subband interaction, the Hartree shift tends to pull down the upper subband away from the

crossing point; and pairing prevails closer to the crossing point which results in the merger of the two subbands into a single paired subband. Following the Maslov and Stone theorem, the conductance in the paired (spin-gapped) phase must be $2e^2/h$ (Ref. 46). We expect that these qualitative predictions are generic for systems with attractive inter-band interactions and not particularly sensitive to the assumptions that we have made: i.e. using the Hartree-Bogoliubov model with local interactions.

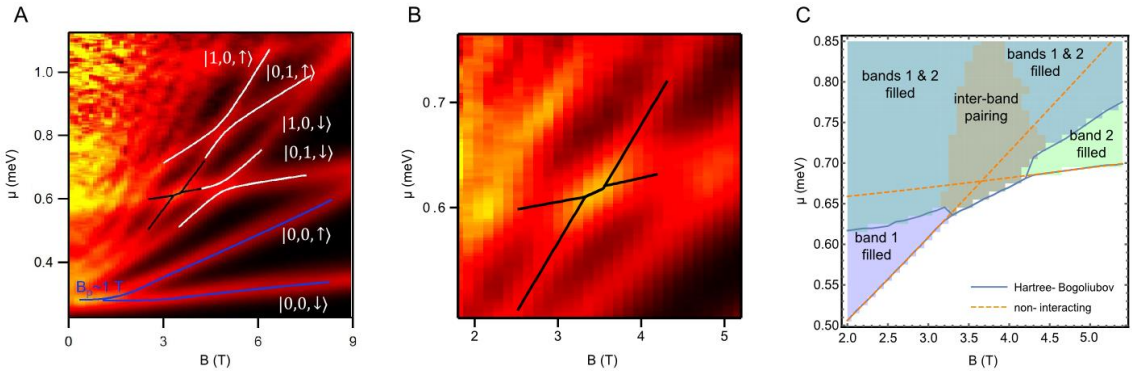

**Figure 5. Electron-electron interactions in electron waveguides.** (**A**) Electron pairing (blue lines), avoided crossing (white lines), and re-entrant pairing (black lines) fittings of transconductance for device A. (**B**) Detailed view of re-entrant pairing data in (A). Here the black lines show the fitting of re-entrant pairing between subband $|0,1,\downarrow\rangle$ and $|1,0,\uparrow\rangle$, with $\Delta_{rp} = 13$ μeV. The white lines are the fitting to the avoided crossing between subband $|0,1,\downarrow\rangle$ and $|1,0,\downarrow\rangle$, with $\Delta_{1,2} = 16$ μeV. (Full set of fitting parameters are listed in Table 1 and Table S2). (**C**) Phase diagram of the Hartree-Bogoliubov model in the $\mu - B$ plane and near the crossing point of $|0,1,\downarrow\rangle$ and $|1,0,\uparrow\rangle$. In producing this diagram, we used the band parameters for device A and set the attractive interaction constants to be $U_H = U_B = 100$ μeV.

**Table 1. Re-entrant pairing fitting parameters for device A and B.**

| Device | Subbands | $k$ (μeV/T) | $b$ (μeV) | $\Delta_{rp}$ (μeV) |
|---|---|---|---|---|
| Device A | $|1,0,\uparrow\rangle$ | 133 | 168 | 13 |
| | $|0,1,\downarrow\rangle$ | 15 | 566 | |
| Device B | $|1,0,\uparrow\rangle$ | 130 | 120 | 10 |
| | $|0,1,\downarrow\rangle$ | 14 | 585 | |

Provided the phase diagram in Figure **5**C, we use a phenomenological model containing the phase boundaries to describe inter-band re-entrant pairing. The basic scenario is when two subbands $E_1 = (k_1 B + b_1)$ and $E_2 = (k_2 B + b_2)$ with opposite spins are tuned closely in energy, they combine as an electron pair, which breaks when the energies are tuned further away. These two subbands would simply cross (orange dashed lines) if there were no electron-electron interaction. In the presence of the attractive pairing interaction, the higher energy subband undergoes an energy shift of $-2\delta_{1(2)}$ so that it can be written as $E'_{1(2)} = k_{1(2)} B + b_{1(2)} - 2\delta_{1(2)}$. And a middle section representing the paired phase emerges. The re-entrant pairing energy $\Delta_{rp}$ can then be extracted: $\Delta_{rp} = \delta_1 + \delta_2$. We are now able to use this model to extract these parameters from the experimental data using the fittings shown in Figure 5C. This process then gives a pairing field range $3.3\text{ T} < B < 3.5\text{ T}$ and a pairing energy $\Delta_{rp} = 13\text{ μeV}$ for subbands $|1,0,\uparrow\rangle$ and $|0,1,\downarrow\rangle$ in device A (see Table 1 for the full fitting parameters).

The observed conductance plateaus are not consistent with a quantum Hall state. The integer quantum Hall effect is defined by an insulating 2D bulk with chiral edge states that are responsible for the quantized conductance. By contrast, LaAlO$_3$/SrTiO$_3$-based electron waveguides lack the insulating bulk region that prevents backscattering. That is to say, the magnetic length ($\ell_B \sim 15$ nm for $B = 3$T) and the confinement length ($\ell_y = 26$ (27) nm for device A (B)) are comparable and no well-defined bulk region is present.

The 3D structure of the electron waveguides is also inconsistent with quantum Hall physics. The cross-section of our waveguides is ellipsoidal with an aspect ratio of 0.5 (vertical/lateral, see Figure 3C, which is well within the 3D regime). This regime is not expected to support stable quantum Hall bilayer states as multiple vertical subbands are occupied. For example, in Figure **5**A, the $|0,0,\uparrow\rangle$ and $|0,1,\downarrow\rangle$ subbands would be unstable and therefore not quantized in a quantum Hall regime, according to Ref. 47. The fact that quantized transport is observed provides further proof that this form of transport is not described by quantum Hall effects.

Finally, the lack of observable quantization at low fields is a consequence of the close spacing of lateral subband modes. The single-particle theory, illustrated in Figure 2E, F, shows that broadening of the subband transitions prevents the individual subbands from becoming resolvable at low magnetic fields; however, they become visible as soon as the magnetic dispersion can clearly separate them in energy. In other waveguides with larger subband spacing, conductance quantization is observable at small magnetic fields (Figure S5).

The observation of quantized conduction in the paired regime ($G = 2e^2/h$ and $|B| < B_p$) signifies that these (non-single-particle) states propagate ballistically, forming an extended state in which electron pairs are bound together while the center-of-mass coordinate remains delocalized. Conduction quantization with steps of $2e^2/h$, rather than $(2e)^2/h$, is consistent with the notion that dissipation takes place not within the channel itself but in the leads, and that electron pairs unbind before they dissipate energy[48-49]. This interpretation is also consistent with the theorem of Maslov and Stone, who argued that the conductance of a Luttinger liquid is determined by the properties of the leads[46]. Specifically, the charge conductance of the channel remains $2e^2/h$ when a spin (i.e. pairing) gap is opened in the channel.

Previous reports of electron pairing in confined 1D structures[42] revealed a range of pairing fields that is consistent with the variation observed in these electron waveguides.

For device A and B, $B_p \approx 1$ T is relatively low compared to $B_p \approx 11$ T in device C. Figure S5 shows additional variation of $B_p$ in two other devices. No specific dependence of $B_p$ on device length can be inferred. Clearly, there are hidden variables that regulate the strength of electron pairing that have yet to be revealed experimentally.

The experiments described here show that electron waveguides provide remarkably detailed insight into the local electronic structure of these oxide interfaces. The level of reproducibility and reconfigurability illustrated by these experiments represents a significant advance in control over electronic transport in a solid-state environment. Correlated electron waveguides offer unique opportunities to investigate the rich physics that is predicted for 1D quantum systems[4]. For example, the number of quantum channels can be tuned to the lowest spin-polarized state (with $G = e^2/h$), forming an ideal spin-polarized Luttinger liquid. The ballistic nature of the transport in 1D is highly surprising, but may be related to the existence of strong electron-electron interactions, which are known to suppress impurity scattering[16-17]. These 1D channels form a convenient and reproducible starting point for emulating a wider class of 1D quantum systems or for creating quantum channels that can be utilized in a quantum computing or quantum simulation platform.

**Materials and Methods**

LaAlO$_3$/SrTiO$_3$ samples are grown by pulsed laser deposition (PLD) under conditions that are described in detail elsewhere (Ref. 31). The electron waveguides are created using c-AFM lithography technique. The wires are written at a tip voltage $V_{tip} = 15$ V except the waveguide, which is created by a two-step voltage sequence. First, we move the AFM tip with $V_{tip} = 8$ V across the LaAlO$_3$ surface to create the main channel.

Next, we repeat the same tip path with a small base voltage ($V_{tip} = 1$ V) and apply two negative voltage pulses ($V_{tip} = -7.5$ V) to create the barriers. The barrier height is determined by the amplitude and duration of the negative pulses. (Note that this method for producing highly transparent barriers is different from Ref. 42, where the c-AFM tip is scanned perpendicular to the nanowire to create tunnel barriers.) Four-terminal transport measurements are carried out at or close to the base temperature of a dilution refrigerator ($T = 50$ mK) and subject to out-of-plane magnetic fields $B$.

**Supporting Information.** Includes finite-bias spectroscopy of electron waveguides, estimation of ballistic scattering length, impact of side gate location on transport, tight-binding Hamiltonian for electron waveguide, description of avoided crossings, measurements on other devices, and measurements on devices with one barrier and no barriers.

**Acknowledgements.** We thank Andrew Daley and Ben Hunt for valuable discussions. This work is supported in part by a Vannevar Bush Faculty Fellowship ONR grant N00014-15-1-2847 (J.L.), DOE DE-SC0014417 (J.L.), AFOSR FA9550-15-1-0334 (C.B.E.), AOARD FA2386-15-1-4046 (C.B.E.), DMR-1629270 (C.B.E.) and 1000 Talents Program for Young Scholars (G.C.).

**Conflict of Interest statement:** The authors declare no competing financial interest.

# Supporting Information

# Quantized Ballistic Transport of Electrons and Electron Pairs in LaAlO$_3$/SrTiO$_3$ Nanowires

Anil Annadi[1,2], Guanglei Cheng[1,2,4], Hyungwoo Lee[3], Jung-Woo Lee[3], Shicheng Lu[1,2], Anthony Tylan-Tyler[1,2], Megan Briggeman[1,2], Michelle Tomczyk[1,2], Mengchen Huang[1,2], David Pekker[1,2], Chang-Beom Eom[3], Patrick Irvin[1,2], Jeremy Levy[1,2*]

[1]Department of Physics and Astronomy, University of Pittsburgh, Pittsburgh, PA 15260, USA.
[2]Pittsburgh Quantum Institute, Pittsburgh, PA, 15260 USA.
[3]Department of Materials Science and Engineering, University of Wisconsin-Madison, Madison, WI 53706, USA.
[4]CAS Key Laboratory of Microscale Magnetic Resonance and Department of Modern Physics, University of Science and Technology of China, Hefei 230026, China
* jlevy@pitt.edu

**Section S1. Finite bias spectroscopy**

Finite-bias spectroscopy is performed through current-voltage (*I-V*) measurements as function of $V_{sg}$ and B to gain more information of the electron waveguides. As shown in Figure S1A, a large finite bias ($V_{sd} \geq V_{sd}^*$) can unevenly populate subbands occupied by oppositely travelling electrons, which gives rise to the so-called half plateaus (1,2). Figure S1B is the finite-bias transconductance plot of device A at $B = 7$ T. The dark regions marked by the numbers are where conductance is quantized. The $0.5e^2/h$ and $1.5e^2/h$ plateaus can be clearly seen in the conductance plot at $V_{sd} = V_{sd}^* = 200$ μV (Figure S1C). The observation of these half plateaus is indicative of very clean transport of the electron waveguide devices, since back scattering is more likely to happen when unoccupied subbands become available at finite biases.

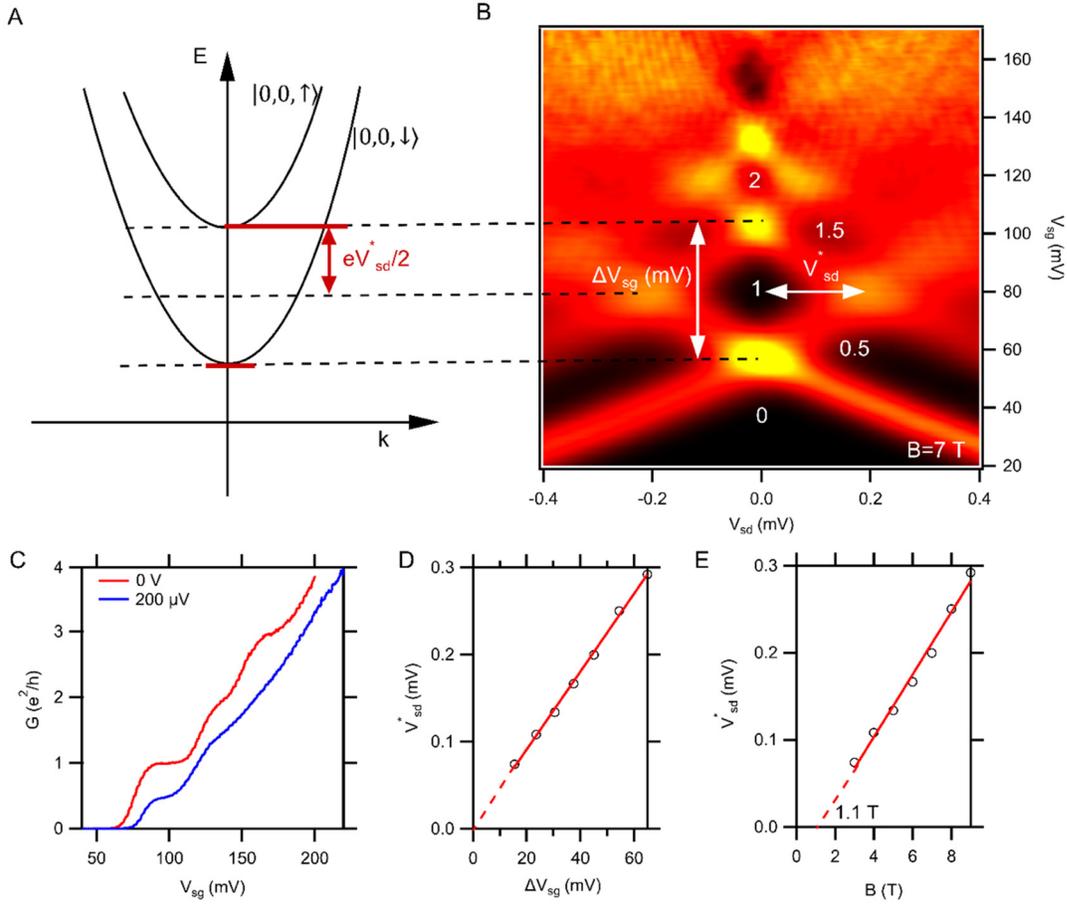

**Figure S1. Finite bias analysis.** (**A**) Illustration of electron occupation of subbands $|0,0,\downarrow\rangle$ and $|0,0,\uparrow\rangle$ at a finite bias $V_{sd}$ in a magnetic field $B = 7$ T. The application of $V_{sd}$ alters the chemical potentials of drain ($\mu_d$) and source ($\mu_s$) to $E_F \pm eV_{sd}/2$. The energy difference $\mu_d - \mu_s = eV_{sd}^*$ (as indicated by two red lines) equals the subband spacing between subbands $|0,0,\downarrow\rangle$ and $|0,0,\uparrow\rangle$. When $V_{sd} < V_{sd}^*$, electrons travelling in opposite directions occupy the same subband $|0,0,\downarrow\rangle$ with conductance quantized to $e^2/h$. When $V_{sd}$ reaches $|V_{sd}^*|$ ($-|V_{sd}^*|$), subband $|0,0,\uparrow\rangle$ becomes available for electrons transmitting from drain (source) and gives rise to half plateau conductance ($1.5\ e^2/h$). (**B**) Transconductance map of device A as a function of $V_{sd}$ and $V_{sg}$ at $B = 7$ T. Each dark band marks the transition between conductance plateaus which are labelled by blue numbers. According to a, the conversion factor $\alpha$ can be extracted through a simple relation $V_{sd}^* = \alpha \Delta V_{sg}$. (**C**) $G$ vs $V_{sg}$ curves of zero bias ($V_{sd} = 0$ V) and finite bias ($V_{sd} = V_{sd}^* = 200$ µV) at $B = 7$ T. Half plateaus are clearly visible at finite bias (blue curve). (**D**) $V_{sd}^*$ dependent on $\Delta V_{sg}$ at magnetic fields from 3 T to 9 T in step of 1 T. The linear relationship and negligible intercept clearly establishes $V_{sd}^* = \alpha \Delta V_{sg}$ with $\alpha = 4.5$ µeV/mV. (**E**) Zeeman splitting between subbands $|0,0,\downarrow\rangle$ and $|0,0,\uparrow\rangle$ with the same field variation in (D). The g factor can be extracted to be $g = 0.6$. Remarkably, subbands $|0,0,\downarrow\rangle$ and $|0,0,\uparrow\rangle$ only split above a critical magnetic field $B_p = 1.1$ T, which is marked by the intercept in the B axis.

Finite-bias spectroscopy is used to extract the lever-arm $\alpha$, which converts gate voltage to chemical potential. As illustrated in Figure S1B, the bright crossing ($V_{sd}^* =$ 200 µV, $V_{sg}$ = 80 mV) marks the transition from one subband to another due to the bias. At this condition, the energy gain induced by the bias $V_{sd}^*$ should equal to subband spacing marked by $\alpha \Delta V_{sg}$ at zero bias, namely $eV_{sd} = \alpha \Delta V_{sg}$. Then $\alpha = eV_{sd}^*/\Delta V_{sg}$ can be precisely extracted by the slope of the $V_{sd}^* - \Delta V_{sg}$ plot at different magnetic fields (Figure S1D). For device A, $\alpha_A$ is found to 4.5 µeV/mV, and the fitted linear curve passes across zero as supposed. Similarly, $\alpha_B = 9.9$ µeV/mV can be extracted for device B, suggesting a stronger coupling of side gate to the waveguide due to the larger size.

The Zeeman splitting between two spin-resolved subbands $|0,0,\uparrow\rangle$ and $|0,0,\downarrow\rangle$ can be used to extract the electron g factor. Figure S1D shows the energy splitting ($eV_{sd}^*$) between these two subbands at various magnetic fields, where spin degeneracy is moved. Then the g factor is given by $g = \frac{eV_{sd}^*}{\mu_B B}$, where $\mu_B$ is the Bohr magneton. And the extracted g factors for device A and B are (within measurement error) the same: $g_{A(B)} = 0.6$.

**Section S2. Estimation of ballistic scattering length**

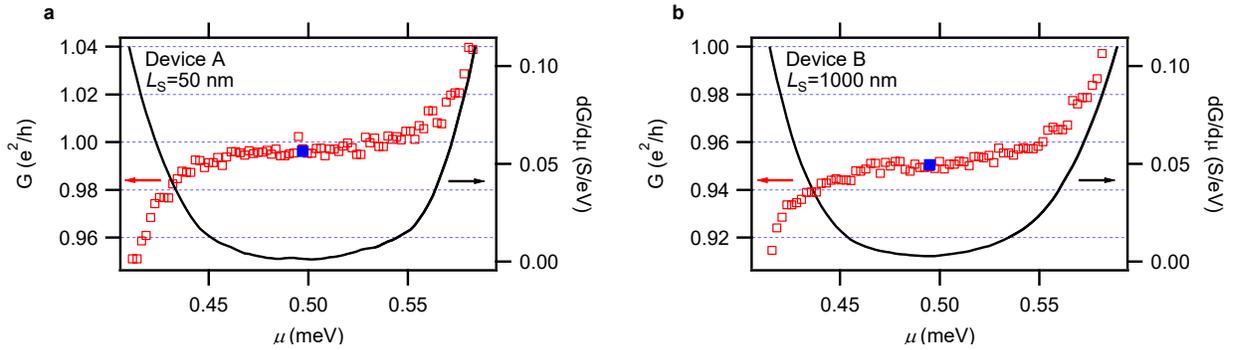

**Figure S2. Quantization of the $1\,e^2/h$ plateau.** (**A**) The first conductance plateau and transconductance for device A. (**B**) The first conductance plateau and transconductance for Device B. For both devices, a fit to the conductance at the minimum of the transconductance is used to estimate scattering lengths.

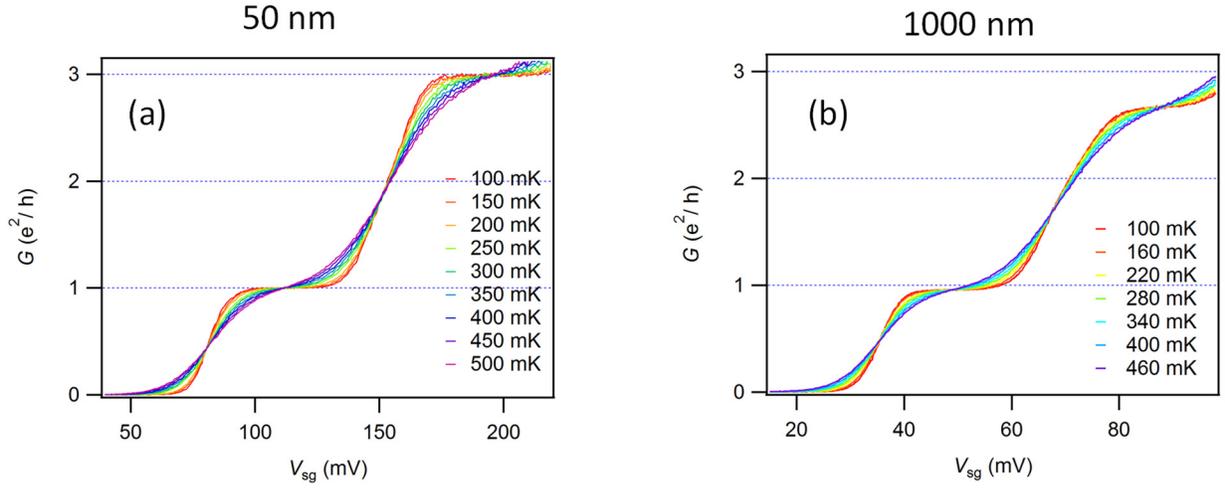

**Figure S3. Temperature dependence of conductance vs gate voltage.** (a) Device A, $L_s$=50 nm. (b) Device B, $L_s$=1000 nm, in an out-of-plane magnetic field of 9 T.

**Table S1. Measured scattering lengths and standard errors for Device A and B**, based on deviations from precise quantization $Ne^2/h$ of conduction.

| Device | Plateau $N$ | $L$ (nm) | $\Delta L$ (nm) | $G(e^2/h)$ | $\Delta G(e^2/h)$ | $L_0$ (µm) | $\delta L_0$ (µm) |
|---|---|---|---|---|---|---|---|
| A | 1 | 50 | 10 | 0.995 | 0.004 | 10 | 8 |
| A | 2 | 50 | 10 | 0.964 | 0.12 | 2 | 14 |
| B | 1 | 1000 | 10 | 0.955 | 0.003 | 21.7 | 1.4 |
| B | 2 | 1000 | 10 | 0.899 | 0.076 | 7.8 | 4.6 |

## Section S3. Impact of side gate location

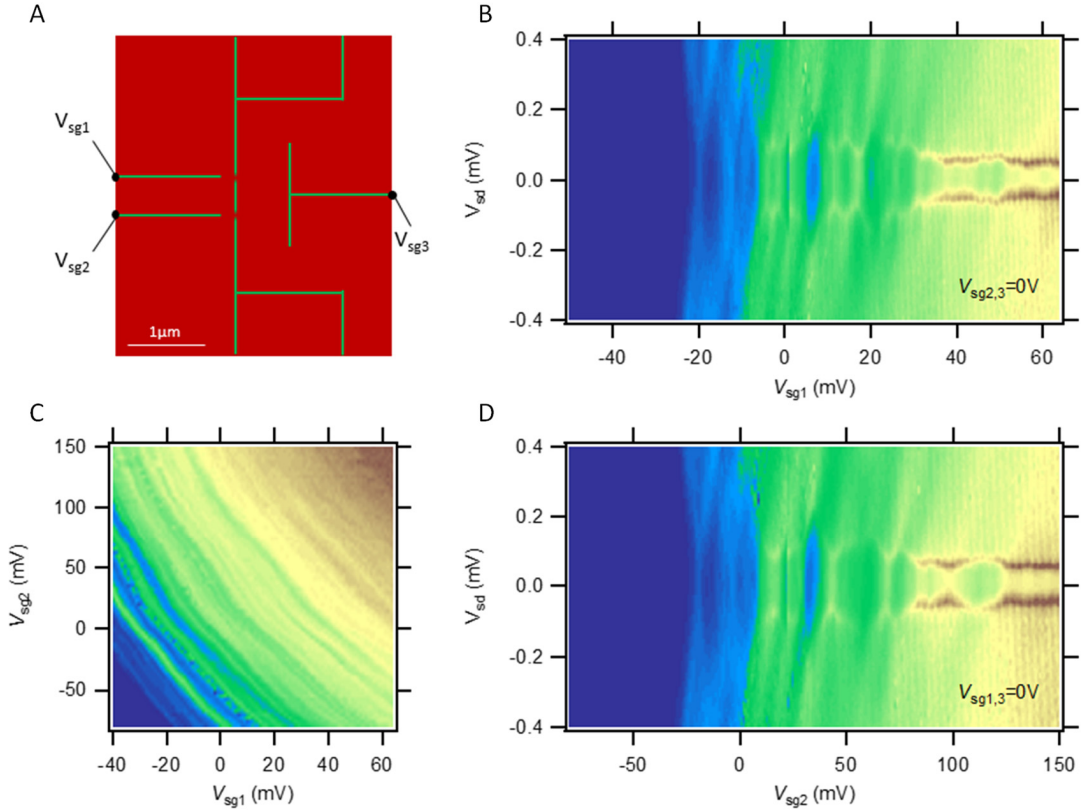

**Figure S4. Effect of gate geometry.** (**A**) Illustration of triple-gate device, consisting of two large tunneling barriers in the main channel, two point-gates near each dot ($V_{sg1}$ and $V_{sg2}$), and a long middle gate to universally tune the device ($V_{sg3}$). (**B**) Conductance measured by lock-in technique as a function of $V_{sg1}$ and $V_{sg2}$, with $V_{sg3}$ floating. While the lever arm for each gate is different, the conductance features exhibit highly similar dependence on gate voltage, regardless of which gate is swept. (**C-D**) Conductance $dI/dV$ as a function of source-drain bias $V_{sd}$ and either $V_{sg1}$ (C) or $V_{sg2}$ (D) while holding the other two gates constant at 0 V.

The physical location of the side gates for LaAlO$_3$/SrTiO$_3$ nanostructures affect the overall lever arm but, somewhat surprisingly, impacts only negligibly the electronic structure within the conducting regions. To illustrate, we show transport results for a single-electron transistor device with multiple side gates (Figure S4A). The differential conductance is shown as a function of four-terminal source-drain voltage $V_{sd}$ and either of the side gates located near one or the other barrier ($V_{sg1}$ or $V_{sg2}$). The results for both gates are nearly identical (Figure S4C, D), which shows that the electric fields are effectively screened and the main result of gating is to change the chemical potential uniformly within the conducting wire segment. The near equivalence of both gates is also shown by plotting the conductance at zero bias versus the two gates (Figure S4B). Differences between the two gates are negligibly small, apart from the factor-of-two difference in lever arms. This insensitivity is likely related to the very large dielectric constant of SrTiO3 at low temperatures.

**Section S4. Tight-binding Hamiltonian for electron waveguide**

As the magnetic field couples to motion in the $xy$-plane, the characteristic length scale and mass in the $y$-direction may be extracted directly form the transconductance data. To extract $l_z$ and $m_z^*$ from $\omega_z$, it is necessary to use a more complete tight-binding model which includes the atomic spin-orbit coupling between the 3 Ti $t_{2g}$ orbitals. The inclusion of this term then allows us to vary the mass $m_z^*$ of the $d_{xy}$ band (and the corresponding masses of the $d_{yz}$ and $d_{zx}$ bands) to see the reduction in the electron $g$ factor (see later discussion in Sec. V). The resulting tight-binding Hamiltonian takes the form

$$H = \sum_{i,j,k} \left[ \sum_{\alpha,s} \left( -t_x^\alpha e^{\frac{i\frac{e}{\hbar}Bjd^2}{\phi_0}} a_{i,j,k}^{\alpha s \dagger} a_{i+1,j,k}^{\alpha s} - t_y^\alpha a_{i,j,k}^{\alpha s \dagger} a_{i,j+1,k}^{\alpha s} - t_z^\alpha a_{i,j,k}^{\alpha s \dagger} a_{i,j,k+1}^{\alpha s} \right) \right.$$

$$+ \frac{\Delta_{aso}}{2} \sum_{s,s'} \left( -i\sigma_y^{ss'} a_{i,j,k}^{d_{xy}s' \dagger} a_{i,j,k}^{d_{yz}s'} + i\sigma_x^{ss'} a_{i,j,k}^{d_{xy}s \dagger} a_{i,j,k}^{d_{xz}s'} \right.$$

$$\left. \left. + i\sigma_z^{ss'} a_{i,j,k}^{d_{yz}s \dagger} a_{i,j,k}^{d_{xz}s'} \right), + h.c. \right] \tag{S1}$$

$$+ \sum_{i,j,k,s,\alpha} \left[ \left( \frac{m_y^{d_{xy}*} \omega_y^2}{2} (jd)^2 + \frac{m_z^{d_{xy}*} \omega_z^2}{2} (kd)^2 + 2t_x^\alpha + 2t_y^\alpha \right.\right.$$

$$\left.\left. + 2t_z^\alpha \right) a_{i,j,k}^{\alpha s \dagger} a_{i,j,k}^{\alpha s} + \frac{g}{2} \mu_B B \sigma_z a_{i,j,k}^{\alpha s \dagger} a_{i,j,k}^{\alpha s} \right]$$

where $t_i^\alpha$ is the hopping in the $i$-direction for the band $\alpha$, $d$ is the lattice constant, $\phi_0$ is the magnetic flux quantum, $\Delta_{aso} = 19.3$meV is the atomic spin-orbit coupling (3), $g$ is the bare-electron $g$ factor, $\mu_B$ is the Bohr magneton, and $a_{i,j,k}^{\alpha s(\dagger)}$ destroys (creates) an electron at site $i,j,k$ with spin $s$ in band $\alpha$. From this, the effective $g$ factor can be extracted and compared to the experimental value to extract $l_z$ and $m_z^*$ from $\omega_z$.

**Section S5. Avoided crossings**

Experimentally, we observe that when two subbands $|n_{y,1}, n_{z,1}, s\rangle$ and $|n_{y,2}, n_{z,2}, s\rangle$ share the same spin quantum number $s$ and are nearly degenerate in energy ($E_{n_{z,1},n_{y,1},s} \approx E_{n_{z,2},n_{y,2},s}$), e.g., $|1,0,\uparrow\rangle$ and $|0,1,\uparrow\rangle$, they form an avoided crossing (Figure 5A). It is tempting to associate avoided crossings with repulsive electron-electron interactions, however the phase diagram of the repulsive version of the model (Eq. (4) does not admit this interpretation). However, an avoided crossing arises naturally if the transverse confinement potential is not separable (4,5). To model these avoided crossings, a simple two-level effective Hamiltonian of the form

$$H_{\text{eff}} = \begin{pmatrix} E_1 & \Delta_{1,2} \\ \Delta_{1,2} & E_2 \end{pmatrix} \quad (S2)$$

is used, where $\Delta_{1,2}$ models the non-separability of the confinement potential by coupling the two states $E_1$ and $E_2$. The chemical potentials at which the two transverse subbands become occupied follows:

$$E_{AV\pm} = \frac{1}{2}(E_1 + E_2) \pm \frac{1}{2}\sqrt{(E_1 - E_2)^2 + 4\Delta_{1,2}^2} \quad (S3)$$

To fit the experimental data, and extract the parameter $\Delta_{1,2}$, we approximate the single particle energy eigenvalue $E_i$ with a linear magnetic field dependence $E_i = k_i B + b_i$ in the vicinity of the avoided crossing (see Figure 5, as well as Table S1).

**Table S2. Avoided crossing fitting parameters for device A and B.**

| Device | Subbands | k (μeV/T) | b (μeV) | $\Delta_{1,2}$ (μeV) |
|---|---|---|---|---|
| Device A | $\|1,0,\uparrow\rangle$ | 167 | 8 | 20 |
| | $\|0,1,\uparrow\rangle$ | 58 | 534 | |
| | $\|1,0,\downarrow\rangle$ | 91 | 195 | 16 |
| | $\|0,1,\downarrow\rangle$ | 15 | 566 | |
| Device B | $\|1,0,\uparrow\rangle$ | 166 | -58 | 34 |
| | $\|0,1,\uparrow\rangle$ | 63 | 529 | |
| | $\|1,0,\downarrow\rangle$ | 95 | 139 | 40 |
| | $\|0,1,\downarrow\rangle$ | 14 | 585 | |

## Section S6. Other devices

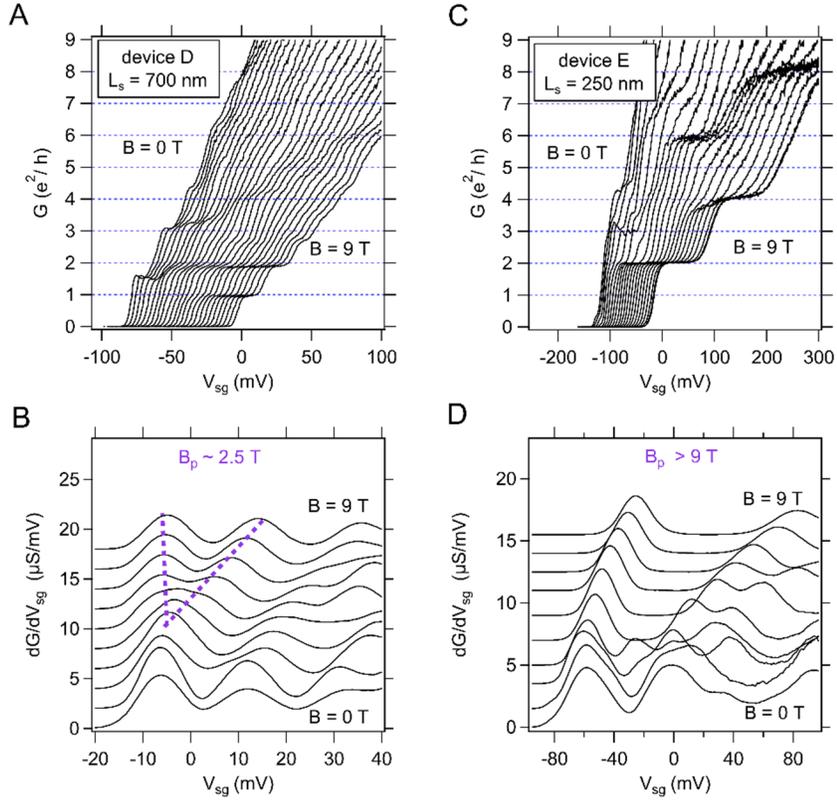

**Figure S5.** Critical magnetic field for splitting the lowest two spin subbands for additional devices D and E (**A** and **C**). Zero-bias conductance $G$ as a function of $V_{sg}$ and $B$ for device D ($L_B = 20$ nm, $L_C = 1500$ nm, $L_s = 700$ nm) and E ($L_B = 20$ nm, $L_C = 500$ nm, $L_s = 250$ nm) fabricated on different samples. Curves are offset for clarity (**B** and **D**). Corresponding transconductance $dG/dV_{sg}$ plots to reveal $B_p$ at which subbands $|0,0,\downarrow\rangle$ and $|0,0,\uparrow\rangle$ start to split. $B_p$ values are high for device D (~2.5 T) and E (>9 T) compared to device A and B in the main text.

## Section S7. Zero-barrier, single-barrier and double-barrier geometry

In GaAs-based heterostructure devices, the number of transverse channels that are transmitted through a quantum point contact is typically controlled by a split top gate. Varying the potential on the split top gate then controls the effective width of the conducting region. In the case of LaAlO$_3$/SrTiO$_3$, similar behavior may be expected in the case where a side gate is used to control a quantum point contact created by a single weak barrier. In Figure S6B, we show the results of varying the $V_{sg}$ for a device with a single barrier in the channel. At all values of the magnetic field, there is no clear quantization of the conductance. This is consistent with the single-particle theory shown in Figs. 2, 3 which holds the width of the conducting channel, $\ell_y$, fixed as $V_{sg}$ is varied. Thus we conclude that, as in the case for the quantum dot geometries used in Ref [41], varying $V_{sg}$ controls the chemical potential of the region between the barriers, as illustrated in Figure 1. When no barriers are present (Figure S6A) there is no observed quantization and the conductance in the nanowire is very large and cannot be tuned to an insulating state.

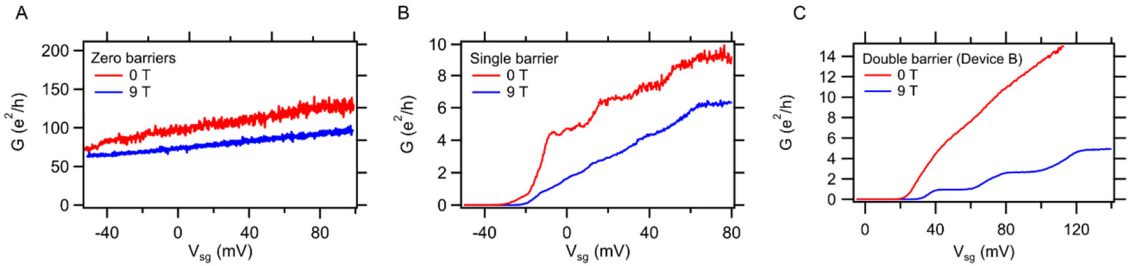

**Figure S6. Control devices with zero and one barriers.** Representative transport data for devices with zero (A) one (B) and two (C) barriers. No conductance quantization is observed in devices with zero and one barrier. Devices with no barriers are not able to be tuned to an insulating state with the applied side gate voltage. Device parameters: (A) $L_c = 80$ nm (B) $L_c = 1800$ nm, $L_b = 20$ nm (C) Device B $L_C = 1.8$ μm, $L_S = 1$ μm, $L_B = 20$ nm.

**Supplementary References**